\documentclass[english,prb,twocolumn]{revtex4-1} 

\usepackage{amsmath}
\usepackage{graphicx} 
\usepackage{bm} 
\usepackage{color} 
\usepackage{amssymb}
\usepackage{verbatim}
\usepackage[colorlinks=true, pdfstartview=FitV, linkcolor=blue, citecolor=blue, urlcolor=blue]{hyperref} 
\usepackage{lineno}

\newcommand{\captionlinespread}{1.05}

\renewcommand{\figurename}{\textbf{Figure}} 
\renewcommand{\thefigure}{\textbf{\arabic{figure}}} 

\def\C        {{$^{13}$C \/}}
\def\N        {{$^{14}$N \/}}
\def\NN       {{$^{15}$N \/}}

\def\ie       {{\it i.e. \/}}

\newcommand{\ee}[1]{\times 10^{#1}}
\newcommand{\mr}[1]{\mathrm{#1}}
\newcommand{\unit}[1]{\,\mathrm{#1}}

\newcommand{\us}{\,\mu{\rm s}}

\newcommand{\Cn}[1]{$^{13}$C$_{#1}$}

\newcommand{\exvalue}[1]{\ensuremath{\left\langle#1\right\rangle}}

\newcommand{\Se}{\hat{S}_e}

\newcommand{\Sz}{\hat{S}_z}
\newcommand{\Ie}{\hat{I}_e}
\newcommand{\Ix}{\hat{I}_x}
\newcommand{\Iy}{\hat{I}_y}
\newcommand{\Iz}{\hat{I}_z}

\newcommand{\aperp}{a_\perp}
\newcommand{\apar}{a_{||}}

\newcommand{\Bo}{B_0}

\newcommand{\dmax}{\delta_\mr{max}}

\newcommand{\eps}{\epsilon}
\newcommand{\fc}{f_\mr{c}}
\newcommand{\fs}{f_\mr{s}}
\newcommand{\Ge}{\Gamma_\mr{e}}
\newcommand{\Gn}{\Gamma_\mr{n}}
\newcommand{\Gbeta}{\Gamma_\beta}
\newcommand{\Gbetasync}{\Gamma_\beta^\mr{sync}}
\newcommand{\Ggamma}{\Gamma_\gamma}
\newcommand{\ms}{m_S}
\newcommand{\phiss}{\phi_\text{ss}}

\newcommand{\SNR}{\mr{SNR}}

\newcommand{\td}{t_\mr{d}}

\newcommand{\treadout}{t_\mr{readout}}
\newcommand{\tbeta}{t_\beta}
\newcommand{\ts}{t_\mr{s}}
\newcommand{\Te}{T_{2,\mr{DD}}}
\newcommand{\Tn}{T_{2,\mr{n}}^\ast}

\newcommand{\avgw}{\langle\omega\rangle}

\newcommand{\wo}{\omega_{0}}

\newcommand{\yn}{\gamma_\mr{n}}

\begin{document}

\title{Tracking the precession of single nuclear spins by weak measurements}
\author{K. S. Cujia}
\author{J. M. Boss}
\author{K. Herb}
\author{J. Zopes}
\author{C. L. Degen\footnote{Email: degenc@ethz.ch}}
%
\affiliation{Department of Physics, ETH Zurich, Otto Stern Weg 1, 8093 Zurich, Switzerland}
\date{\today}
%
%


\maketitle



\textbf{
Nuclear magnetic resonance (NMR) spectroscopy is a powerful technique for analyzing the structure and function of molecules, and for performing three-dimensional imaging of the spin density. At the heart of NMR spectrometers is the detection of electromagnetic radiation, in the form of a free induction decay (FID) signal \cite{hahn50}, generated by nuclei precessing around an applied magnetic field.  While conventional NMR requires signals from $10^{12}$ or more nuclei, recent advances in sensitive magnetometry \cite{poggio10,wrachtrup16} have dramatically lowered this number to a level where few or even individual nuclear spins can be detected
\cite{jelezko02,mamin13,staudacher13,loretz14apl,muller14}.
It is natural to ask whether continuous FID detection can still be applied at the single spin level, or whether quantum back-action modifies or even suppresses the NMR response.
Here we report on tracking of single nuclear spin precession using periodic weak measurements
\cite{korotkov01,clerk10,gross15,gefen18}.
Our experimental system consists of \C nuclear spins in diamond that are weakly interacting with the electronic spin of a nearby nitrogen-vacancy center, acting as an optically readable meter qubit.  We observe and minimize two important effects of quantum back-action: measurement-induced decoherence \cite{colangelo17} and frequency synchronization with the sampling clock \cite{shiga12,jordan05}.
We use periodic weak measurements to demonstrate sensitive, high-resolution NMR spectroscopy of multiple nuclear spins with \textit{a priori} unknown frequencies.  Our method may provide the optimum route for performing single-molecule NMR
\cite{degen09,ajoy15,perunicic16}
at atomic resolution.
}


Measurement back-action, an important feature of quantum measurements
\cite{weber14,blok14},
can usually be neglected in NMR because the spin-detector coupling is extremely weak.  One prominent exception is radiation damping \cite{bloembergen54}, where the collective coupling of the nuclear ensemble gives rise to a damping of the magnetic resonance by the electric detection circuit.  As nuclear ensembles become smaller, eventually consisting of only few or even a single nuclear spin,  the close coupling to the detector is expected to modify
\cite{sidles92,bergh96}
or inhibit \cite{berman05}
the free evolution of the spin.  Recent work on ensembles of cold atoms \cite{smith06} and trapped ions \cite{colangelo17} reported simultaneous tracking of spin angle and amplitude through the use of weak, quantum-non-demolition measurements, indicating an avenue for mitigating back-action.  Here, we show that it is possible to track the precession of a single nuclear spin
and to extract the two central pieces of information in NMR: the free precession frequency and the dephasing time.

\begin{figure*}[t]
    \centering
    \includegraphics[width=0.85\textwidth]{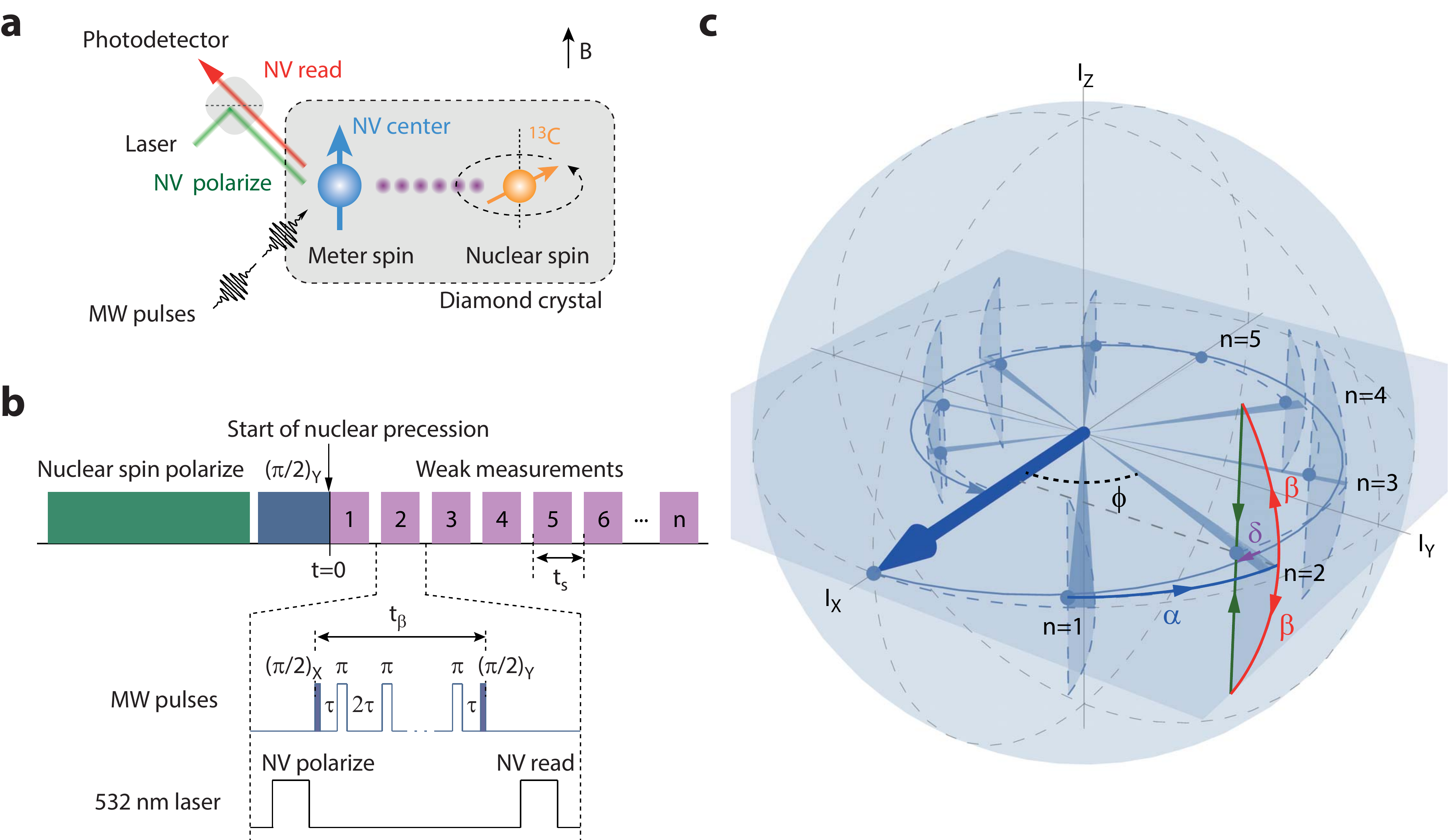}
    \caption{
		\linespread{\captionlinespread}\selectfont{}
{\bf Scheme for tracking of single nuclear spin precession.}
{\bf a}, Measurement system:  A meter qubit, implemented by the electronic spin of a single NV center in diamond, weakly probes a precessing \C nuclear spin.  The meter is read out via a strong optical measurement.  The interaction between electronic and nuclear spins can be turned on and off by microwave manipulation of the electronic spin.
{\bf b}, Measurement protocol:  The nuclear spin is polarized into $\Iz$ (green) and rotated by $\pi/2$ (blue) to initiate precession.  Weak measurements (purple) are implemented by a conditional rotation of the electronic-nuclear spin pair followed by an optical readout of the NV spin component $\Sz$.  We implement the conditional rotation as a decoupling sequence of periodically spaced $\pi$ pulses (inset) where $\tau\sim \pi/(2\wo)$.  To track the nuclear precession, weak measurements are repeated at instances of the sampling time $\ts$.
{\bf c}, Bloch-sphere representation of the nuclear state evolution.  At $t=0$ the nuclear vector points along the +X axis (bold arrow).  A Z-oriented magnetic field $\Bo$ drives spin precession in the XY plane.  Each sampling interval consists of (i) an advance of the spin angle $\phi$ by $\alpha=\wo\ts$ (modulo $2\pi$) due to free precession (blue arrow), a conditional rotation around the $\pm$X axes by $\beta$ (red arrows), and a strong readout of the electronic meter qubit causing projection of the nuclear spin onto the XY plane (green arrows).  Each weak measurement reduces the amplitude of the $\Iy$ component by $\cos(\beta)$ and gives a phase kick $\delta$ (purple arrow) to the spin angle.  Periodic weak measurements of the nuclear spin produce an inward spiraling precession trace.
    }
    \label{fig:fig1}
\end{figure*}


To probe the coherent precession of a single nuclear spin we implemented the measurement system depicted in Fig. \ref{fig:fig1}a.  Our system consists of a \C nucleus (spin $I=1/2$) isolated in the nearly spin-free lattice of a diamond crystal.  The nuclear spin undergoes a free precession around the Z axis with an angular velocity given by the Larmor frequency $\wo = \yn B_0$, where $B_0$ is the local magnetic field and $\yn$ the nuclear gyromagnetic ratio.  To detect the nuclear precession, we periodically couple the nuclear spin to the electronic spin of a nearby NV center acting as an optically readable meter qubit.  We monitor the precession by probing the nuclear $\Ix$ spin component by means of a conditional rotation via the interaction Hamiltonian $\hat{H}_\mr{meas} = g\,2\Ix\Sz$, where $g$ is a coupling constant.  For two-spin systems, this Hamiltonian can be realized, for instance, by a Carr-Purcell-type dynamical decoupling sequence applied to the meter spin
\cite{taminiau14,boss16}.
The interaction with the nuclear spin imprints a signal proportional to $\sin(\beta)\langle\Ix\rangle$ onto the optically measurable $\Sz$ spin component of the meter, where $\beta=g\tbeta$ is the rotation angle that determines the measurement strength.
Importantly, by varying the interaction time $\tbeta$, we can smoothly tune the measurement strength, or turn it off completely.  This feature allows us to explore the cross-over from the strong ($\beta\approx \pi/2$) to the weak ($\beta\rightarrow 0$) measurement regime.

In addition to providing the signal, the interaction with the meter also modifies the trajectory of the nuclear evolution, leading to back-action.  The state evolution is illustrated in Fig. \ref{fig:fig1}c and can be sequenced into two unitary rotations and a partial projection: During each sampling interval, (i) the nuclear spin angle $\phi$ accumulates a free precession phase $\alpha = \wo\ts$, where $\ts$ is the dwell time.  (ii) The interaction with the meter qubit rotates the nuclear spin around the X axis by the angle $\pm\beta$, where the sign is conditional on the $\Sz$ state of the meter. (iii) Projective optical readout of the meter $\Sz$ component collapses the nuclear vector onto the XY plane, which reduces the amplitude of the $\Iy$ component by a factor of $\cos(\beta)$ and gives a small phase kickback $\delta \approx -\frac14\beta^2\sin(2\phi)$ to the spin angle $\phi$.  Under precession, $\Ix$ and $\Iy$ alternate roles as the measured and perturbed variables, leading to an exponential decay of the spin amplitude with a decay rate $\Gbeta \approx \beta^2/(4\ts)$ and an average precession rate $\avgw:=\langle d\phi/dt\rangle \approx \wo$.  By plotting the meter output as a function of readout index $n$, we therefore expect a decaying oscillation, allowing us to extract estimates for the precession frequency $\wo$ and the decay rate $\Gamma$.  Importantly, by making the measurement weak (small $\beta$), we can quadratically suppress measurement-induced decoherence\cite{gefen18} $\Gbeta\propto\beta^2$ while only linearly reducing the signal amplitude $A\propto\beta$.


\begin{figure*}[t!]
    \centering
    \includegraphics[width=0.85\textwidth]{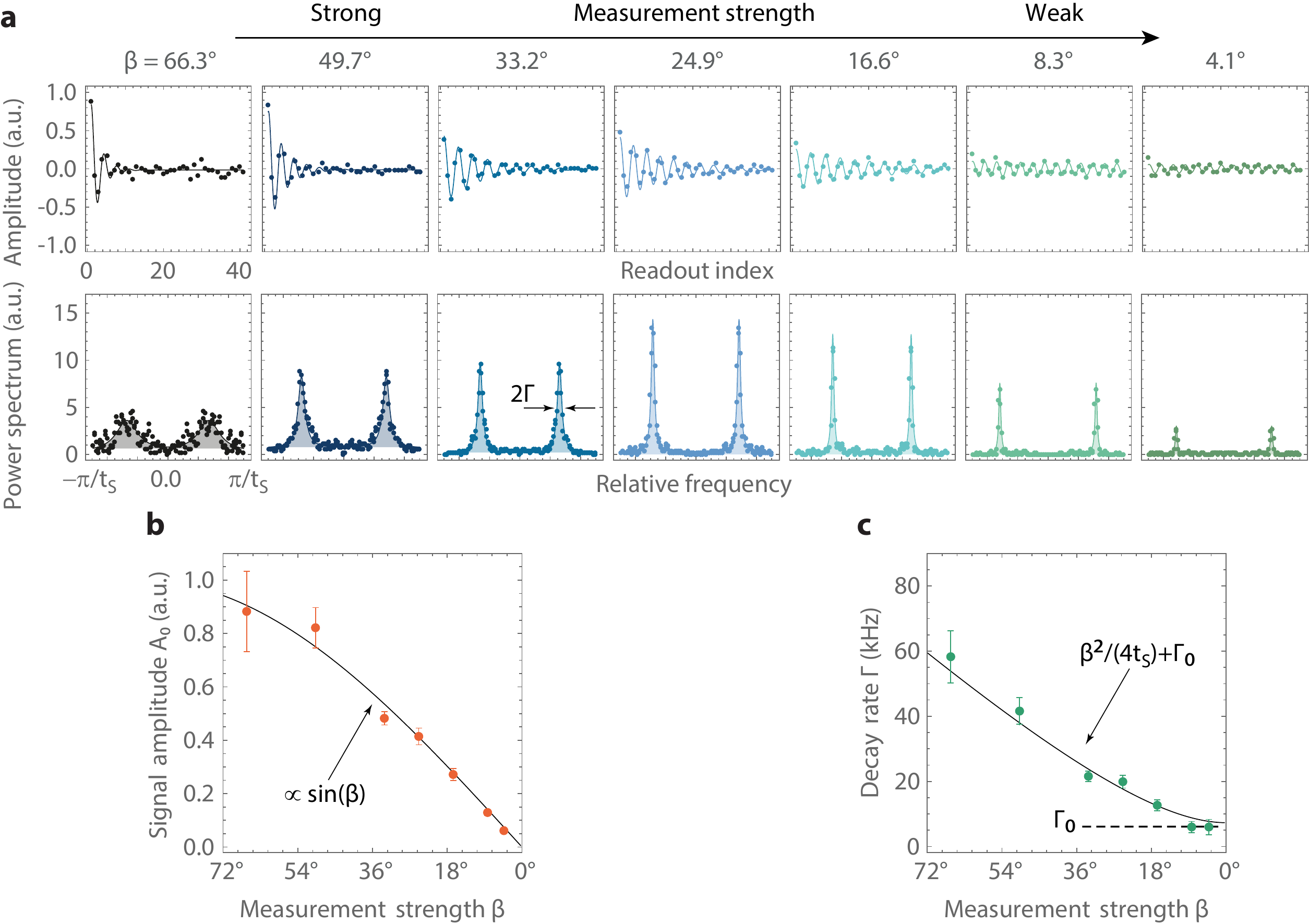}
    \caption{
		\linespread{\captionlinespread}\selectfont{}
{\bf Experimental observation of single \C precession.}
{\bf a}, Time traces (upper panels) and corresponding power spectra (lower panels) for decreasing measurement strengths $\beta$.  We vary $\beta$ by adjusting the interaction time $\tbeta$ from $3.9$ to $0.25\unit{\us}$.  The dots show the experimental data and the solid lines are fits to exponentially decaying sinusoids (time traces) and Lorentzians (power spectra).  The spectra are obtained from undersampled time traces and are plotted on a relative frequency scale.  Time traces are averaged over $0.7\ldots 3.8\ee{6}$ repetitions of the experiment \cite{supplementary}.
{\bf b}, Signal amplitudes $A_0$ as a function of measurement strength $\beta$. The dots and error bars depict the fit results from ({\bf a}).
The solid curve is the function $\sin(\beta)$ that is vertically scaled to fit the data.
{\bf c}, Decay rates $\Gamma$ as a function $\beta$. The dots and error bars depict the fit results from ({\bf a}).  The solid curve is the function $\Gamma(\beta,\ts)=\beta^2/(4\ts) + \Gn$, where $\Gn$ accounts for the intrinsic nuclear dephasing with a decay time $\Tn=(\Gn)^{-1} = 134\unit{\us}$.
    }
    \label{fig:fig2}
\end{figure*}

Experimental traces of single \C spin precession are shown in Fig. \ref{fig:fig2}.  Seven experiments are plotted where the measurement strength $\beta$ decreases from left to right.  We control $\beta=g\tbeta$ through the interaction time $\tbeta$, where the coupling $g = 2\pi\cdot 46.8\unit{kHz} \approx \aperp/\pi$ is set by the transverse hyperfine coupling parameter $\aperp$ between the electronic and nuclear spin \cite{boss16}. $\aperp$ is determined by an independent calibration (Methods).  For all measurements, the nuclear spin is initially polarized into the $\Iz$ state by repetitive initialization and rotated into $\Ix$ to initiate precession (Methods).  The nuclear evolution is then tracked by periodically probing $\Ix$ at instances of a sampling time $\ts$.  The measurement output is the photon count of the NV meter spin, averaged over $\sim 10^6$ repetitions of the experiment to reach adequate SNR.

The experiment of Fig. \ref{fig:fig2}a clearly demonstrates that the free precession of a single nuclear spin can be continuously tracked, and that we can minimize back-action by simply reducing the measurement strength $\beta$:  When probing $\Ix$ strongly ($\beta = 66.3^\circ$), the oscillation collapses within a few measurement cycles.  Conversely, when using weak measurements ($\beta=4.1^\circ$), the oscillation persists over the entire measurement record with little decay.  By fitting each trace to an exponentially decaying oscillation, we can extract the signal amplitudes and decay rates (Fig. \ref{fig:fig2}c,d).  As expected, for small $\beta$ the signal amplitude $A_0 \propto \sin\beta \approx \beta$ is proportional to $\beta$ while the decay rate $\Gamma = \beta^2/(4\ts) + \Gn$ scales with $\beta^2$.  The offset $\Gn$ represents the intrinsic dephasing of the nuclear spin, $\Gn = (\Tn)^{-1}$, mainly caused in our experiments either by a residual hyperfine interaction to the NV center or by slow drifts in the static magnetic bias field (Methods).

\begin{figure*}[t]
    \centering
    \includegraphics[width=0.99\textwidth]{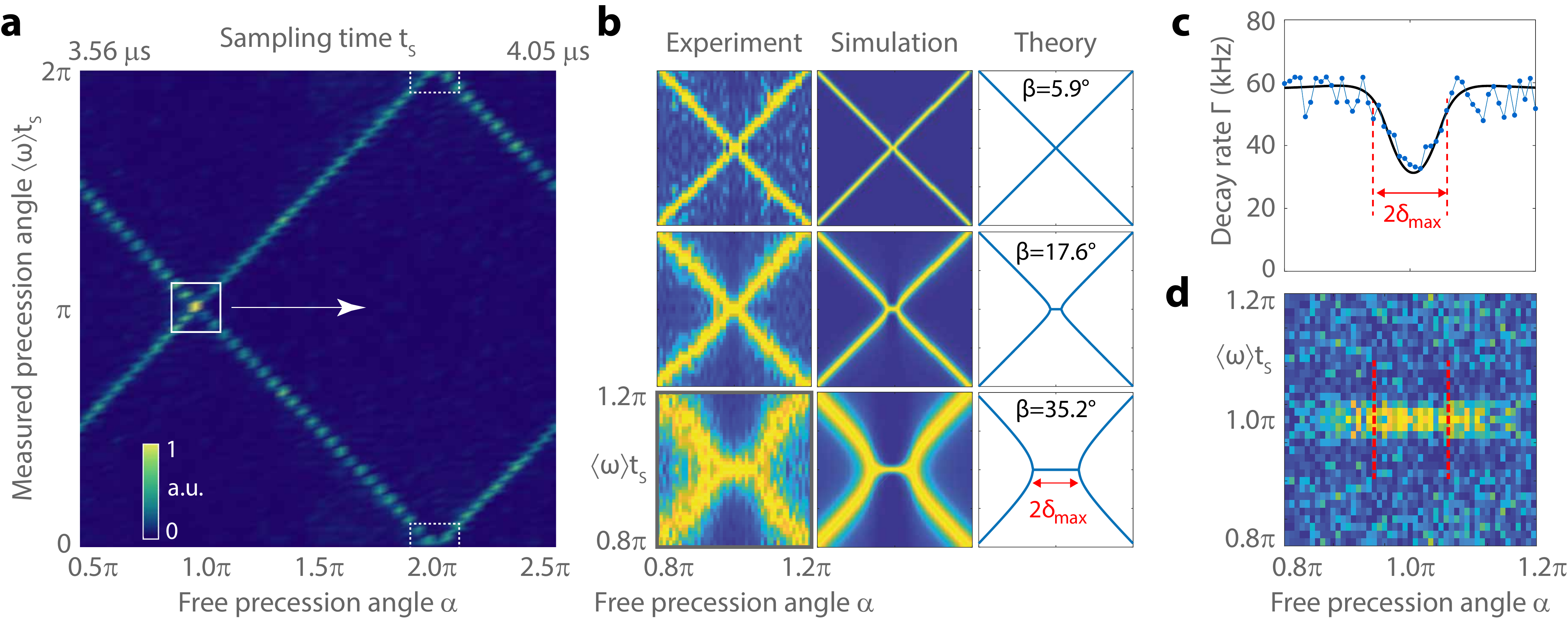}
    \caption{
		\linespread{\captionlinespread}\selectfont{}
{\bf Frequency synchronization.}
{\bf a}, Normalized power spectra (vertical axis) plotted as a function of the free precession angle $\alpha$ (horizontal axis).  We vary $\alpha$ by incrementing the sampling time from $3.56\unit{\us}$ to $4.05\unit{\us}$ in steps of $10\unit{ns}$.  The peak positions in the power spectra identify the average precession angle $\avgw\ts$.  
{\bf b}, Zoom into the region $\alpha=0.8...1.2\pi$ showing frequency synchronization.  The locking range $2\dmax = \beta^2/2$ increases with the square of the measurement strength $\beta$ (top to bottom).  The left panels show the experimental observation, the middle panels show a density matrix simulation, and the right panels show the theoretical expectation (Methods).
{\bf c}, Nuclear dephasing rate $\Gamma$ plotted as a function of $\alpha$, revealing suppressed dephasing under synchronization.
{\bf d}, Spontaneous alignment of the nuclear spin.  The plot shows the same measurement as in ({\bf b}), bottom row, recorded without prior initialization of the nuclear spin.  Under synchronization, the nuclear spin spontaneously aligns with $\Ix$ leading to a bright signal peak.  No peak is observed in the non-synchronized regime.
The measurement strength in ({\bf a}) is $\beta=8.6^\circ$.  The measurement strength in ({\bf c,d}) is $\beta = 35^\circ$ and the locking range is $2\dmax = 0.06\pi$.
    }
    \label{fig:fig3}
\end{figure*}

We next address the question of whether the continuous weak observation of the nuclear spin alters the free precession frequency, which is the most important quantity in NMR spectroscopy.  Our experimental estimate for the precession frequency $\avgw$ is the time derivative of the spin angle averaged over one time trace.  We extract $\avgw$ by fitting the peak position in the power spectrum (Fig. \ref{fig:fig2}a).  To analyze whether $\avgw$ corresponds to the ``true'' free precession frequency $\wo$, we record a set of time traces for varying forward precession angle $\alpha = (0.5...2.5)\pi$.  We tune $\alpha=\wo\ts$ by incrementing the sampling time $\ts$ over one Larmor period.
Fig. \ref{fig:fig3}a shows the resulting power spectra (vertical axis) as a function of precession angle $\alpha$ (horizontal axis) in a normalized color plot.  Clearly, the peaks appear at the diagonal positions where $\avgw\ts = \alpha$.  A statistical analysis of the peak frequencies confirms that $\avgw\ts$ and $\alpha$ agree within the experimental error (Extended Data Fig. \ref{fig:linearfreqfit_fig2res}).  Fig. \ref{fig:fig3}a demonstrates that, in general, weak measurements do not modify the free precession frequency.

Although Fig. \ref{fig:fig3}a suggests that $\avgw\ts = \alpha$ for all values of $\alpha$, a closer inspection reveals that the equality is violated near the cross-over points where $\alpha = k\pi; \; k\in\mathbb{N}$ is a multiple of $\pi$ (white boxes).  As $\alpha\rightarrow k\pi$ (Fig. \ref{fig:fig3}b), the precession rate $\avgw$ abruptly locks to the fixed value of $\pi/\ts$, meaning that the precession synchronizes with the sampling clock.  This synchronization becomes more pronounced as the measurement strength is increased (Fig. \ref{fig:fig3}b, top to bottom).
The phenomenon can be explained by the phase kickback $\delta$ of the weak measurement (purple arrow in Fig. \ref{fig:fig1}c): Because $\Iy$ is scaled while $\Ix$ is not, the spin vector is effectively squeezed towards the X axis.  Once the forward precession angle becomes smaller than the maximum kickback, $\mod(\alpha,\pi) < \dmax = \beta^2/4$, the spin vector is trapped and synchronization sets in.  This basic explanation is supported by the excellent agreement of the theoretical description (Methods) with our experimental observations and complementary density matrix simulations (Fig. \ref{fig:fig3}c).

Frequency synchronization is accompanied by several intriguing features which have been the subject of extensive theoretical work
\cite{korotkov01,jordan05,jordan06}.
First, the phase locking stabilizes the precession \cite{jordan05} and suppresses dephasing of the nuclear spin (Fig. \ref{fig:fig3}c).  This suppression is strongest for $\alpha = k\pi$ since the spin vector is always close to the X axis, and less effective once $\alpha$ moves away from multiples of $\pi$.  Second, the synchronization leads to a spontaneous initialization into $\pm\Ix$ from a random spin orientation
\cite{liu17}
(Fig. \ref{fig:fig3}d).  Both features could be enhanced by active feedback control
\cite{jordan06,gillett10}
of $\alpha$, via either the sampling period $\ts$ or the precession frequency $\wo$.  Note that while our observed frequency synchronization is reminiscent of the quantum Zeno effect \cite{korotkov01,kalb16}, it is more closely related to the concepts of spin locking \cite{slichter90} and atomic phase locking
\cite{shiga12,kohlhaas15},
because projective measurement is not required.


\begin{figure}[hp]
    \centering
    \includegraphics[width=0.99\columnwidth]{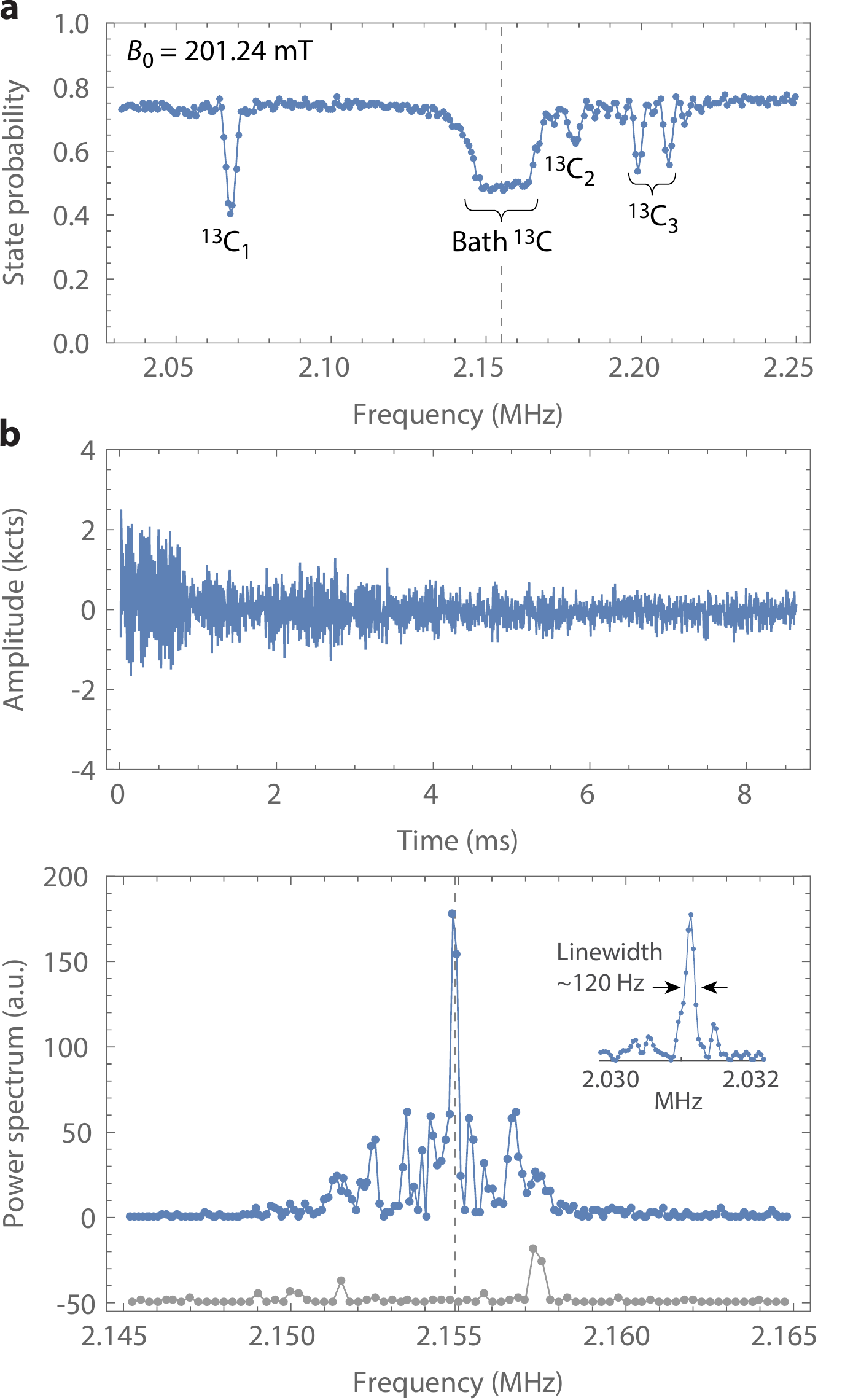}
    \caption{
        \linespread{\captionlinespread}\selectfont{}
{\bf High-resolution NMR spectroscopy of weakly coupled \C nuclei}.  
{\bf a}, Conventional dynamical decoupling (DD) spectrum \cite{taminiau12}.
Sharp dips reflect three individual carbon nuclei (\Cn{1-3}) that are strongly coupled to the NV center.  The broad central dip is due to the distant \C spin bath.  The vertical dashed line indicates the nuclear Zeeman frequency of $2.1549\unit{MHz}$.  The DD sequence uses interaction times $\tbeta = 222\ldots 246\unit{\us}$ with $N=200$ pulses.
The nuclei are not pre-polarized.
{\bf b}, Weak measurement time trace (upper panel) and power spectrum (lower panel, blue dots) of the \C bath for the same NV center.  The time trace is obtained from $n=1,520$ readouts measured at $\ts=5.68\unit{\us}$ dwell.  The interaction time is $\tbeta = 1.86\unit{\us}$.  The gray data show a nuclear Ramsey spectrum \cite{maurer12,abobeih18} for direct comparison, vertically offset for clarity.  The spectra are normalized to one standard deviation of the respective baseline noise.  The pre-polarization time is $36\unit{ms}$ for both spectra (Methods). 
The inset shows a separate high-resolution spectrum of a single, distinct \C at a bias field $\Bo=190.2\unit{mT}$.
Averaging times are 24 hours in (a) and 9.5 hours in (b).  No post-processing is applied beyond an overall scaling and baseline subtraction.
}
    \label{fig:fig4}
\end{figure}

We finally demonstrate that periodic weak measurements are ideally suited for applications in nanoscale NMR spectroscopy \cite{mamin13,staudacher13}, because they enable simultaneous detection of multiple nuclear spins with \textit{a priori} unknown frequencies at a competitive signal-to-noise ratio.
We illustrate this capability by performing \C NMR of a specific NV center's ($T_1=2.3\unit{ms},\,\Te=216\unit{\us}$) carbon environment.  Initial characterization by dynamical decoupling (DD) spectroscopy
\cite{taminiau12,kolkowitz12,zhao12}
(Fig. \ref{fig:fig4}a) indicates the presence of three nearby \C nuclei coupled to this NV center, evidenced by distinct dips in the spectrum.  The central broad dip is due to the distant bath of weakly coupled \C spins.  The bath dip is not further resolved because meter relaxation limits the spectral resolution of DD spectroscopy to $\Te^{-1}$ (here $\sim 4.3\unit{kHz}$, see Methods).
\cite{biercuk11,degen17}
By contrast, weak measurement spectroscopy is not restricted by meter relaxation, but only by the duration of the time trace $n\ts$ and by nuclear dephasing.
\cite{schmitt17,boss17,glenn18},
Consequently, the spectral resolution can become very high.
Fig. \ref{fig:fig4}b shows that weak measurement spectroscopy is capable of resolving rich structural detail of the \C bath.
Many sharp peaks are observed that correspond to individual or small clusters of \C.  Importantly, detected nuclei have small coupling constants $g/(2\pi)$ of a few kHz, comparable to or below the meter decoherence rate $\Te^{-1}$ \cite{gefen18}.  These nuclei are difficult to access by conventional DD, correlation \cite{laraoui13} or nuclear Ramsey spectroscopies
\cite{maurer12,abobeih18}
(gray dots in Fig. \ref{fig:fig4}b).  Further advantages of weak measurement spectroscopy include a wide receiver bandwidth (Methods and Extended Data Figs. \ref{fig:wm_bw_nspins} and \ref{fig:sensitivity_comparison}) and the ability to selectively detect weak signals despite the presence of strongly coupled nuclei (Methods).


Although demonstrated on carbon nuclei inside the diamond host crystal, our experiments can be readily extended to outside molecules by using diamond chips with near surface NV centers
\cite{mamin13,staudacher13,loretz14apl}.
For these applications, much broader spectral peaks are expected.  Future experiments will therefore likely require the use of advanced solid-state NMR techniques.
For example, homo- and heteronuclear decoupling pulses could be interspersed with weak measurements to suppress spin-spin interactions \cite{slichter90,maurer12}.  Alternatively, spin dilution and isotope labeling of the target molecule could be employed \cite{loquet11}.  For molecular imaging applications, single molecules could be isolated in a spin-free host material and analyzed by three-dimensional localization spectroscopy
\cite{zopes18ncomms,sasaki18,zopes18prl}.
Although many technical issues still remain, the prospect for three-dimensional imaging of single molecules with atomic resolution and chemical selectivity provides a strong motivation for exploring these possibilities.


\vspace{0.5cm}



{\bf Acknowledgments:}
This work has been supported by the Swiss National Science Foundation through project grants 200020\_156100, 200020\_175600 and through the NCCR QSIT, and by the European Commission through DIADEMS grant 611143 and ASTERIQS grant 820394.
We thank Renbao Liu, Alex Retzker and Tim Taminiau for helpful discussions, Kevin Chang for experimental support, and Marius Palm for proof-reading the manuscript.\\

{\bf Author contributions:}
C.L.D. conceived the project. K.S.C., J.M.B. and K.H. carried out the experiments with the support of J.Z and analyzed the data.  K.S.C., C.L.D. and J.M.B. performed the simulation and theoretical analysis of weak measurements.  All authors discussed the results and participated in writing the manuscript.\\




\vspace{0.5cm}
{\bf Methods}\\

\textbf{Diamond samples:}
Two single-crystal diamond chips were used for experiments.
Sample A (NV1, NV3-7) was an electronic-grade, natural abundance (1.1\% \C) diamond membrane.  We etched nano-pillars \cite{babinec10,momenzadeh15} into the membrane surface to increase the photon collection efficiency.
Sample B (NV2) was an unstructured diamond chip overgrown with $20\unit{nm}$ of enriched $^{12}$C (99.99\%), $1\unit{nm}$ of enriched $^{13}$C (estimated in-grown concentration $\sim$5-10\%), and a $5\unit{nm}$ cap of enriched $^{12}$C (99.99\%); further details on the sample are given in Ref. \onlinecite{unden18} (``Sample B'').  
NV centers in both samples  were created by $^{15}$N$^+$ ion implantation at an energy of $5\unit{keV}$ and subsequent annealing at 850$^{\circ}$~C.  We chose the \NN species to discriminate implanted NV centers from native (\N) NV centers.  Both samples were cleaned in a 1:1:1 mixture of H$_2$SO$_4$:HNO$_3$:HClO$_4$ and baked at $465^{\circ}$~C before mounting them in the setup.  The continuous wave (CW) photon count rate was $250-700\unit{kC/s}$ for sample A and $40-50\unit{kC/s}$ for sample B.\\

\textbf{Experimental setup:}
Experiments were performed with a custom-built confocal microscope equipped with a green $532\unit{nm}$ excitation laser and a $630-800\unit{nm}$ detection path using a single photon detector.  Optical pulses were generated by an acousto-optic modulator and gating of arriving photons was realized by time-tagging and software binning of photon counts.
Microwave pulses for manipulating the electronic spin were synthesized using an arbitrary waveform generator (AWG, Tektronix AWG5012C), up-converted to $\sim 2.5\unit{GHz}$ using a local oscillator (Hittite HMCT2100) and a quadrature mixer (IQ1545, Marki microwave) and subsequently amplified (Gigatronics GT-1000A). The pulses were delivered to the NV center using a coplanar waveguide defined on a quartz cover slip by photolithography. The transmission line was terminated on an external 50$\,\Omega$ load.
Radio-frequency pulses for manipulating nuclear spins were synthesized using an AWG (National Instruments PCI-5421) and subsequently amplified (Mini-Circuits LZY-22+). The pulses were transmitted using a planar micro-coil \cite{zopes18prl}. The micro-coil had a 3dB-bandwidth of $19\unit{MHz}$.  The \C Rabi frequency was $\sim 25\unit{kHz}$.

We used a cylindrical permanent magnet to create a bias field $B_0\sim 200\unit{mT}$ at the NV center location.  We aligned $B_0$ with the NV symmetry axis by adjusting the relative location of the permanent magnet.  The alignment was optimized by fitting to a set of EPR lines recorded at different magnet locations and by maximizing the CW photon count rate.\\

\textbf{Tracking of magnetic field drifts:}
For long acquisitions (several hours), the magnetic bias field drifted by typically a few Gauss, leading to variations in the EPR frequency of a few MHz and variations in the \C Larmor frequency of a few kHz.  These drifts were likely caused by a temperature-related change of the magnetization of the permanent magnet providing the bias field \cite{ma02}.  We continuously tracked and logged the field drift during a measurement via the EPR resonance of the NV center and adjusted the microwave excitation frequency in real time.  For some spectra (Fig. \ref{fig:fig4}b, inset) we applied a post-correction \cite{rosskopf17} to compensate for the drift in the \C spectrum.  The residual drift was on the order of 30-50\,ppm.\\

\textbf{Detection protocol:}
A detailed diagram of the detection protocol is given in Extended Data Fig. \ref{fig:extended_pulse_diagram}.  The protocol consisted of three steps: (i) Polarization of the nuclear spin, (ii) a $90^\circ$ pulse to initiate the free nuclear precession, and (iii) a series of $n$ weak measurements.

\textit{Polarization:}
Two methods were used to polarize nuclear spins.  For the experiments shown in Figs. \ref{fig:fig2} and \ref{fig:fig3} we followed the method of Ref. \onlinecite{taminiau14}.  We first initialized the electronic spin into $\ms=0$ using a laser pulse.  We then applied a conditional nuclear $\pi/2$ X rotation, implemented as a resonant Carr-Purcell-Meiboom-Gill (CPMG) decoupling sequence applied to the electronic spin.  We next applied a nuclear $\pi/2$ Z rotation, implemented as a waiting time of duration $\tau$.  We then applied a second conditional nuclear $\pi/2$ X rotation, implemented by another CPMG sequence with the initial and final electronic $\pi/2$ pulses omitted.  We finally applied a laser pulse to repolarize the electronic spin into $\ms=0$.  We typically repeated this sequence a few times until the polarization saturated (Extended Data Fig. \ref{fig:repetitive_initialization}).
For the experiments shown in Figs. \ref{fig:fig4} and Extended Data Figs. \ref{fig:wm_dd_comparsion}, \ref{fig:wm_check} we used an amplitude-ramped NOVEL sequence \cite{london13}. We first polarized the resident \NN nuclear using two c-NOT gates on the electronic and \NN spins, respectively \cite{rosskopf17}. The c-NOT gates were implemented by a selective $\pi$ pulse on the electron spin (conditional on the state of the \NN spin), followed by a selective $\pi$ pulse on the \NN spin (conditional to the $m_S=0$ state of the electron spin), and a final laser pulse.  We then initialized the electronic spin into the $\ms=0$ state and subsequently applied the NOVEL sequence, consisting of an electronic $\pi/2$ rotation followed by a $30\unit{\us}$ long spin-lock pulse that was phase-shifted by $90^\circ$.  The amplitude of the spin lock was ramped by typically $10\%$ around the resonance condition.  We repeated this sequence (without the \NN initialization) up to $1200$ times, in order to make the total polarization time $t_\mr{polarize}$ similar to the duration of the weak measurement trace.  Extended Data Fig. \ref{fig:wm_check} shows weak measurement spectra as a function of initialization time.

\textit{Nuclear $90^\circ$ pulse:}  For the experiments shown in Figs. \ref{fig:fig2} and \ref{fig:fig3}, the nuclear $\pi/2$ rotation was applied by means of another resonant CPMG sequence, again with the initial and final electronic $\pi/2$ pulses omitted. For the experiments shown in Figs. \ref{fig:fig4} and Extended Data Figs. \ref{fig:wm_dd_comparsion}, \ref{fig:wm_check} the nuclear $\pi/2$ was directly applied via the external micro-coil.  The duration of the $\pi/2$ pulse was typically $10\unit{\us}$.

\textit{Weak measurements:} The weak measurements were implemented by polarizing the electronic spin into $\ms=0$ using a laser pulse, applying a resonant CPMG sequence of duration $\tbeta$, and reading out the electronic state with a second laser pulse.  To obtain a meter output that is proportional to $\Ix$, the initial and final electronic $\pi/2$ pulses of the CPMG sequence must be phase shifted by $90^\circ$ with respect to each other \cite{taylor08,boss17,schmitt17,glenn18}.\\

\textbf{Calibration of electronic and nuclear spin parameters:}
The resonance frequency of the electronic spin was calibrated by an optically-detected magnetic resonance spectroscopy scan. 
The nuclear Zeeman frequency as well as the parallel and transverse hyperfine coupling parameters ($\apar$ and $\aperp$) were calibrated using correlation spectroscopy \cite{boss16}.  The reported coupling constants $g$ are equal to $\pi/(2\tbeta)$ where $\tbeta$ is the CPMG duration leading to a $\beta=\pi/2$ rotation.  All calibration data are provided as Supplementary Information.\\

\textbf{Derivation of the measurement-induced decay rate:}
Assume that after $n$ weak measurements the spin vector has length $r_n<1$.  Then, the length $r_{n+1}$ after the $(n+1)$'th weak measurement is
$r_{n+1} = \sqrt{1-\sin\phi^2(1-\cos\beta^2)} r_n$, where $\phi$ is the instantaneous spin angle just before the measurement and $\beta$ is the angle of the conditional rotation.  If $\beta\ll\pi/2$ is small, we find $r_{n+1} \approx \cos(\beta\sin\phi) r_n$.  For a series of $n$ weak measurements at spin angles $\phi$ uniformly distributed between 0 and $2\pi$, $r_n \approx \langle \cos(\beta\sin\phi) \rangle^n \approx \exp(-n\beta^2/4) \approx \exp(-\Gbeta t)$, where the average is over $\phi=0...2\pi$ and where
\begin{align}
\Gbeta = \beta^2/(4\ts)
\label{eq:gphi}
\end{align}
is the exponential decay rate.  A quantum mechanical derivation of Eq. (\ref{eq:gphi}) is given in Supplementary Note 1.\\

\textbf{Derivation of frequency synchronization:}
The phase kick $\delta$ is given by the difference of the spin angle before and after a weak measurement.  Using the simple geometric picture of rotations on the Bloch sphere (Fig. \ref{fig:fig1}c) we find that $\delta = \arctan(\tan\phi\cos\beta) - \phi$.
Reforming this expression into $\tan(\phi+\delta) = \tan\phi\cos\beta$ and applying the tangent's sum rule, we find
\begin{equation}
\tan\delta = -\frac{\tan\phi(1-\cos\beta)}{1+\tan\phi^2\cos\beta} \approx - \frac{\beta^2}{4}\sin 2\phi
\label{eq:freqsync}
\end{equation}
where the approximation holds for small $\beta\ll\pi/2$.  Eq. (\ref{eq:freqsync}) shows that: (i) the phase kick is independent of the sign of $\beta$, thus both projections of the electronic spin generate the same kick, (ii) the phase kick alternates direction on each quadrant since $\sin(2\phi)$ changes sign every $90^\circ$, (iii) frequency synchronization is therefore expected when the free precession angle $\alpha$ is an integer multiple of $\pi$, and (iv) the maximum phase kick is $|\dmax| \approx \beta^2/4$, defining the locking range for synchronization.
Eq. (\ref{eq:freqsync}) further determines the steady-state spin angle $\phiss$ under synchronization, found by setting $\delta = -\alpha$ (modulo $\pi$) and solving for $\phi$:
\begin{equation}
\phiss = \frac12\sin^{-1}\left[\frac{4\tan\alpha}{\beta^2}\right] \approx \frac{2\alpha}{\beta^2}
\label{eq:phiss}
\end{equation}
where the last expression is for small $\phiss\ll\pi/4$.

Frequency synchronization suppresses nuclear dephasing both because measurement-induced decoherence is reduced, and because Z phase noise is canceled by the phase kickback $\delta$.  The measurement-induced decoherence is reduced because the spin vector is always close to $\Ix$, which is not disturbed by the weak measurement.  For reasonably small $\phiss\ll\pi/4$, the decoherence rate is $\Gbetasync \approx 2\Gbeta\sin^2\phiss \approx \Gbeta \alpha^2/(2\delta^2)$, where $\Gbeta$ is given by Eq. (\ref{eq:gphi}) and $\alpha$ is modulo $\pi$.  Clearly, $\Gbetasync$ vanishes as $\alpha\rightarrow 0$.  Frequency synchronization is also expected to suppress the intrinsic nuclear dephasing due to Z phase noise, however, we did not observe this effect in the present study.

We finally calculate the average precession frequency $\avgw$ under weak continuous measurements for all values of the precession angle $\alpha$.  Inside the locking range,
\begin{align}
\avgw = \frac{k\pi}{\ts} \quad\quad \text{if $\mod(\alpha,\pi) \leq \dmax$}
\end{align}
where $k\in\mathbb{N}$ is determined by the undersampling.  Outside the locking range, the instantaneous precession frequency at spin angle $\phi$ is $\omega(\phi) = (\alpha + \delta)/\ts = \wo - \frac{\beta^2}{4\ts}\sin 2\phi = \wo - \Gbeta\sin 2\phi$.  The experimentally measured precession frequency $\langle\omega\rangle$ is the time average of $\omega(\phi)$.  We calculate $\langle\omega\rangle$ by integrating over one full precession cycle $T$,
\begin{align}
\langle\omega\rangle &= \frac{2\pi}{T}
= 2\pi\left[\int_0^{2\pi}\frac{d\phi}{\omega(\phi)}\right]^{-1} \\
&= \wo' \sqrt{1-(\Gbeta/\wo')^2}
\quad\quad\text{if $\mod(\alpha,\pi)>\dmax$} \nonumber
\label{eq:avgw}
\end{align}
where $\wo'$ is equal to $\wo$ modulo $k\pi/\ts$.  Eq. (\ref{eq:avgw}) shows that $\avgw$ diverges with the square of $\alpha\propto\wo'$ as
$|\alpha|\rightarrow\dmax$,
where $\alpha$ is again modulo $\pi$.
A quantum mechanical derivation is given in Supplementary Note 1.\\

\textbf{Intrinsic nuclear dephasing:} 
Contributing factors to the intrinsic nuclear dephasing rate $\Gamma_0 = (\Tn)^{-1}$ include:

\textit{Dipolar broadening:} The dipolar coupling between \C nuclei causes a homogeneous broadening of the nuclear magnetic resonance.  This effect is expected to be $<50\unit{Hz}$ in our system \cite{maurer12,abobeih18}.

\textit{Magnetic field drift:} Slow drifts in the static magnetic bias field cause fluctuations in the Larmor frequency, as discussed above.  With frequency tracking, the contribution is on the order of 30-50\,ppm, equivalent to $50-100\unit{Hz}$.

\textit{Residual hyperfine interaction during optical illumination:} 
During optical readout and re-initialization of the NV center via a non-resonant laser pulses, the NV center cycles through its electronic states until reaching the $\ms = 0$ spin polarized steady state.  During this process, the NV center stochastically jumps back and forth between spin states and possibly electronic charge states.  This leads to random Z rotations of the nuclear spin due to the secular part of the hyperfine interaction $\apar\,2\Sz\Iz$, where $\apar$ is the parallel hyperfine coupling constant \cite{kalb16}.  The random Z angle is $\gamma \propto \apar\treadout$, where $\treadout$ is the duration during which $\ms\neq 0$.  The length of the spin vector after $n$ weak measurements is approximately
$\langle\cos\gamma\rangle^n = \exp(-n\langle\gamma^2\rangle/2) = \exp(-\Ggamma t)$,
where $\Ggamma \propto \apar^2\treadout^2/(2\ts)$.
For the measurements shown in Fig. \ref{fig:fig2}, we obtain $\Ggamma = 7.4\unit{kHz} = (134\unit{\us})^{-1}$ using $\treadout = 2.8\unit{\us}$ and the experimental parameters given in Supplementary Data 1.
For the weakly coupled \C shown in the inset of Fig. \ref{fig:fig4}b, we obtain $\Ggamma = 3.2\unit{Hz} = (310\unit{ms})^{-1}$.\\

\textbf{Sensitivity:}
As derived in Supplementary Text 2, the signal-to-noise ratio (SNR) per unit time for an optimum choice of parameters ($\Gbeta^{-1} \approx \Tn \approx n\ts \approx t_\mr{polarize} \gg \Te$, $\tbeta \approx \ts$) scales as 
\begin{align}
\SNR \propto \left\{
\begin{array}{ll}
	\eps\sqrt{C_0/\Tn}  & \quad\text{for $g>(\Tn\Te)^{-\frac12}$}  \\
	\eps\sqrt{C_0 g^2\Te} & \quad\text{for $g<(\Tn\Te)^{-\frac12}$}
\end{array} 
\right. 
\end{align}
where $\epsilon$ is the optical readout contrast, $C_0$ the number of photons collected per NV spin readout, $\Te$ the NV spin coherence time and $\Tn$ the nuclear dephasing time.  Importantly, weak measurements allow maintaining an approximately constant SNR for couplings $g$ less than the sensor decoherence rate $\Te^{-1}$.
For the experiment in Fig. \ref{fig:fig4}, $\epsilon \approx 0.35$, $C_0 \approx 0.1$, $\Te \approx 200\unit{\us}$ and $\Tn \approx 5\unit{ms}$.\\

\textbf{Receiver bandwidth:}
The detection bandwidth of weak measurement spectroscopy is determined by the filter function \cite{biercuk11} of the CPMG DD sequence (see Fig. \ref{fig:fig1}b).  The filter function is according to Eq. (66) of Ref.~\onlinecite{degen17},
\begin{equation}
W(f) = \frac{\sin(\pi f \tbeta)}{\pi f \tbeta} [1-\sec(2\pi f\tau)]
\label{eq:ddfilter}
\end{equation}
where we choose $2\tau \sim \pi/\wo$. (Note that the interpulse delay is $2\tau$ in our manuscript, see Fig. \ref{fig:fig1}b).  The bandwidth (full width at half maximum, FWHM) of this filter function is approximately $\tbeta^{-1}$, see Ref.~\onlinecite{degen17}, Eq. (69).  Because the interaction time $\tbeta$ is typically short, the bandwidth is correspondingly wide -- an important advantage of weak measurement spectroscopy.  For the experiment in Fig. \ref{fig:fig4}b, $\tbeta^{-1} \approx (1.86\unit{\us})^{-1} \approx 538\unit{kHz}$.  The Nyquist bandwidth of the spectrum is smaller, because $\ts>\tbeta$.  Therefore, signals outside the Nyquist rate will be aliased and folded back into the spectrum.  The folding can be corrected for using compressive sampling techniques \cite{boss17}.\\
%

\textbf{Selectivity to weak signals:}
Because signals from strongly coupled nuclei are strongly attenuated by measurement back-action, they do not contribute to the spectrum.  Weak measurement spectroscopy therefore enables selective detection of weak signals in the presence of strong couplings.
The selectivity can be tuned by the interaction time $\tbeta$, which determines the angle $\beta=g\tbeta$ and therefore the peak amplitude and linewidth of the power spectrum (see Fig. \ref{fig:fig2}a).  Maximum signal intensity results for $\Gbeta\approx\Gn$, yielding the optimum time $\tbeta$ for a chosen coupling $g$,
\begin{equation}
\tbeta \approx \left( \frac{4\Gn\ts}{g^2} \right)^{1/2} 
\end{equation}
For Fig. \ref{fig:fig4}b, $\Gn \approx 100\unit{Hz}$, $\tbeta = 1.86\unit{\us}$ and $\ts = 5.68\unit{\us}$, corresponding to an optimum coupling parameter $g/(2\pi) \approx 4.1\unit{kHz}$.  
To sensitively detect nuclei over a range of coupling values, a series of spectra with appropriate $\tbeta$ values should therefore be collected, as shown in Extended Data Fig. \ref{fig:wm_check}a.\\

\textbf{Dynamical decoupling spectroscopy:}
The spectrum shown in Fig.~\ref{fig:fig4}a was acquired by sweeping the interpulse delay $2\tau$ of a CPMG DD sequence \cite{taminiau12}.  At each point, we first applied a laser pulse to initialize the electronic spin into the $\ms=0$ state, applied a CPMG sequence of $\pi$ pulses with XY8 phase cycling, and read out the final spin state by a second laser pulse.  The nuclear spins were not polarized and the CMPG sequence used equal phases (X or --X) for the initial and final $\pi/2$ pulses.  The data shown in Fig.~\ref{fig:fig4}a were recorded at the $5^\mr{th}$ harmonic order \cite{taminiau12} and the reported frequency is $f=5/(4\tau)$.  Further experimental parameters are given in Supplementary Data 1.\\

\textbf{Data analysis:}
The data in Fig.~\ref{fig:fig2}a were fitted by $A(n) = A_0 e^{-\Gamma_n n}\sin(2\pi f_0 n + \phi_0)$ with $A_0$, $\Gamma_n = \Gamma\ts$, $f_0$, and $\phi_0$ as free fit parameters (time traces) and by $S(f) = S_0\Gamma^2/[(f^2-f_0^2)+\Gamma^2] + S_1$ with $\Gamma$, $f_0$, $S_0$ and $S_1$ as free fit parameters (power spectra).  The time traces were zero-padded $4\times$ before calculating the power spectra.  The data in Fig.~\ref{fig:fig2}b were fitted by $A_0(\beta) = a\sin(\beta)$, with $a$ as a free fit parameter. The data in Fig.~\ref{fig:fig2}c were fitted by $\Gamma(\beta,\ts) = \frac{\beta^2}{4\ts}+\Gamma_0$, with $\Gamma_0$ as a free fit parameter.
The decay rates in Fig.~\ref{fig:fig3}c were extracted from the data plotted in Fig.\ref{fig:fig3}b (bottom) using the same fit functions as in Fig.~\ref{fig:fig2}a.
%
The spectra shown in Fig.~\ref{fig:fig4}b were obtained from undersampled time traces, and unfolded to the correct frequency using the center frequency $\fc=1/(4\tau)$ of the filter function and the sampling frequency $\fs=1/\ts$.  Both spectra are normalized to one standard deviation of the baseline noise.  The spectrum shown in the inset was zero padded $4\times$ before applying the Fourier transformation.  The reported linewidth is the FWHM of the power spectrum, and was extracted by fitting the spectrum to the function $S(f) = (1-S_0)\Gamma^2/[(f-f_0)^2+\Gamma^2] + S_0$.\\

\textbf{Density matrix simulations:}
To verify our theoretical description for the experimental results, we performed density matrix simulations of the coupled electron-nuclear two-spin system.  We initialized the density matrix into the $\sigma = (\Sz+\Se)\otimes(\Ie+\Ix)$ state and calculated the evolution under $n$ weak measurements. For each weak measurement, we applied a unitary evolution under a CPMG sequence, traced out the electronic spin states leaving only the nuclear spin part $\sigma_I$ of the density matrix, and reinitialized the system into $\sigma = (\Sz+\Se)\otimes\sigma_I$.  The value plotted in Fig.~\ref{fig:fig3}b is the expectation value $\Sz$ of the electronic spin and we verified that $\Sz\propto\Ix$.  The simulations used the experimental parameters listed in Supplementary Data 1 as an input.  We did not account for finite pulse durations or spin relaxation in the simulations.\\

%
%




\renewcommand{\figurename}{\textbf{Extended Data Figure}} 
\renewcommand{\thefigure}{\textbf{\arabic{figure}}} 
\setcounter{figure}{0} 


\begin{figure*}[h!]
    \centering
    \includegraphics[width=\textwidth]{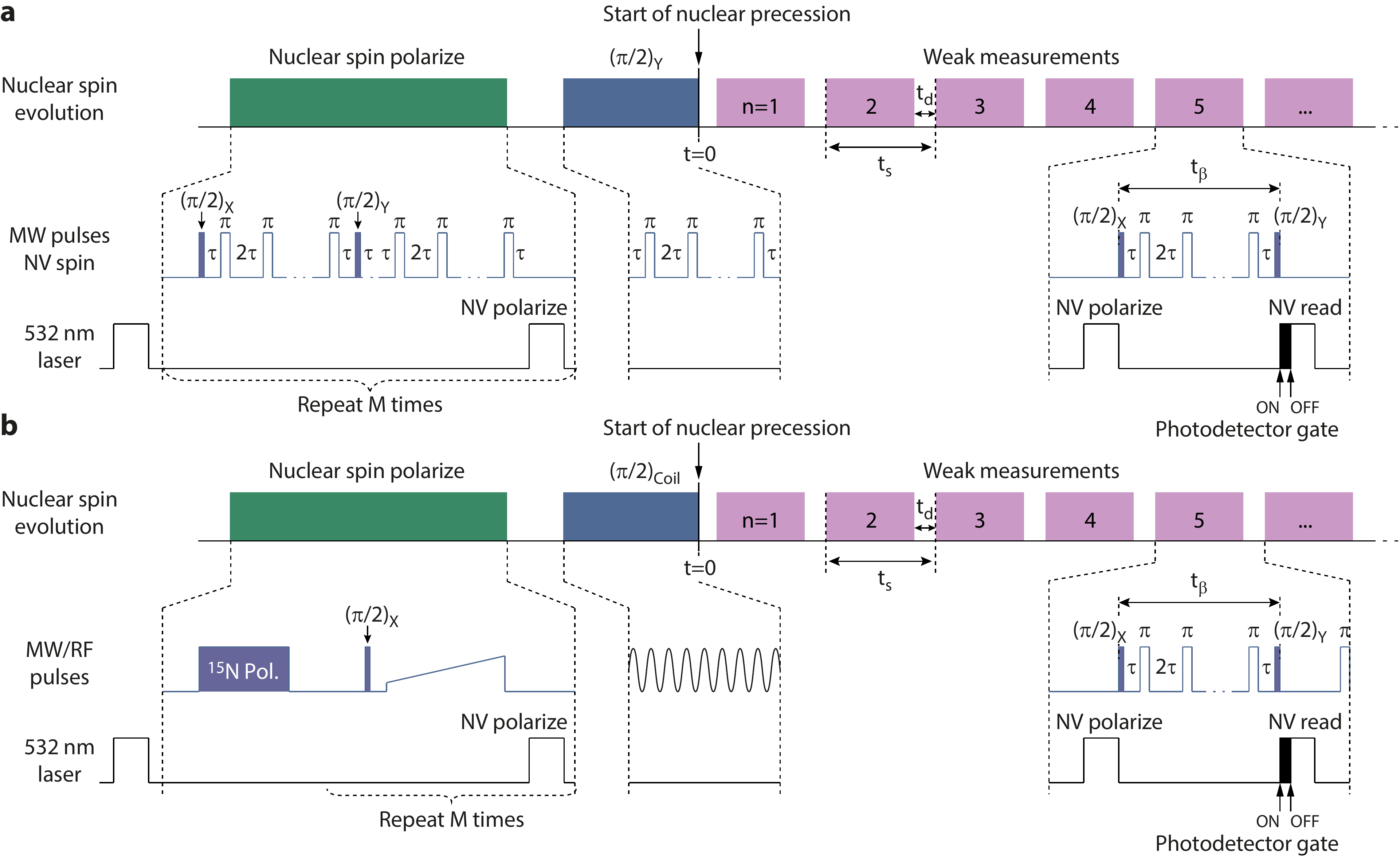}
    \caption{
		\linespread{\captionlinespread}\selectfont{}
    {\bf Extended pulse-timing diagram.}
        {\bf a}, Protocol used for Figs.~2 and 3. The sensor is initially polarized by means of a $\sim 532\unit{nm}$ laser pulse. We subsequently apply a polarization transfer gate \cite{taminiau14} in a repetitive fashion. From the nuclear spin perspective, the polarization sequence consists of a $\pi/2$ conditional X-rotation, followed by a $\pi/2$ Z-rotation and a subsequent $\pi/2$ conditional X-rotation. We implement the conditional X-rotations as a resonant Carr-Purcell-Meiboom-Gill (CPMG) decoupling sequence applied on the electronic spin. The CPMG sequences consist of a periodic train of microwave $\pi$ pulses with alternating phases and an interpulse delay of $2\tau$. The CPMG sequence is resonant when the interpulse delay matches the effective Larmor frequency of the nuclear spin, $\tau \approx \pi/(2(\yn\Bo+0.5\apar))$, where $\apar$ is the parallel dipolar hyperfine coupling between the sensor and the nuclear spin. This condition leads to an effective interaction between the sensor and nuclear spin of the form $g\,2\hat{S}_z\hat{I}_x$ \cite{boss16}, \ie, a simultaneous conditional rotation, where $g \approx \aperp/\pi$ is a coupling rate determined by the transverse dipolar hyperfine coupling between the sensor and the nuclear spin. We implemented Z-rotations as a waiting time of duration $\tau_0=\pi/(2\yn\Bo)$ if the sensor was polarized into $m_S=0$, or of duration $\tau$ if it was not. Z-rotations can alternatively be implemented as non-resonant CPMG sequences, which effectively decouple the evolution of the nuclear spin from the sensor \cite{taminiau14}.
        To initiate precession, we apply a $\pi/2$ Y-rotation on the nuclear spin, implemented as another $\pi/2$ X-rotation followed by a $\pi/2$ Z-rotation. We then probed the nuclear state X-projection, $\exvalue{\hat{I}_x}$, at intervals of a sampling time $\ts$ by means of weak measurements. Each weak measurement instance was implemented as a resonant CPMG decoupling sequence of duration $\tbeta$, sandwiched between two $\pi/2$ pulses whose axes were orthogonal, here X and Y. We used a laser pulse to readout the sensor $\Sz$ state upon each weak measurement instance. An additional delay time $\td$ was used to adjust the sampling time $\ts$. 
%
        {\bf b}, Protocol used for Fig.~\ref{fig:fig4} and Extended Data Figs. \ref{fig:wm_dd_comparsion} and \ref{fig:wm_check}. These experiments probed a bath of \C spins whose hyperfine couplings were not known \textit{a priori} nor directly measured, and the nuclear spins were directly manipulated by means of an external rf coil.  We first polarized the host \NN spin using c-NOT gates on the electronic and \NN spin, implemented as a selective $\pi$ pulse on the electron spin (conditional on the state of the \NN spin), followed by a selective $\pi$ pulse on the \NN spin (conditional to the $m_S=0$ state of the electron spin), and a final laser pulse.
        Subsequently, we applied a NOVEL polarization transfer sequence consisting of a $\pi/2$ X-rotation on the electron spin followed by a linear-ramp spin-lock pulse along its Y axis. The relative amplitude increment of the spin-lock pulse was typically 10\% around the resonant amplitude value and the duration $30\unit{\us}$ leading to a bandwidth of $\approx 100\unit{kHz}$. This procedure was repeated up to M=1,200 times.  To initiate precession, we applied a $\pi/2$ pulse on the bath of \C nuclear spins using the rf coil.  We additionally included a $\pi$ pulse on the electron spin during the delay time $\td$ in order to recover information about the dipolar couplings of the bath spins.
    }
		\label{fig:extended_pulse_diagram}
\end{figure*}

\begin{figure*}[h!]
    \centering
    \includegraphics[width=0.5\textwidth]{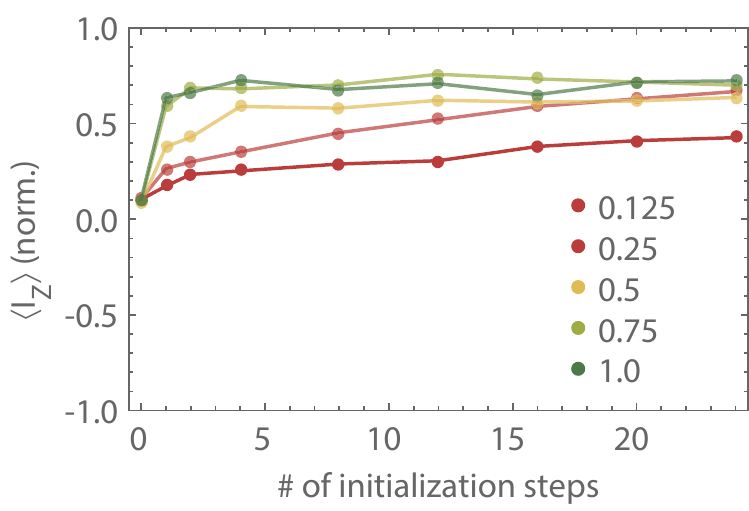}
    \caption{
		\linespread{\captionlinespread}\selectfont{}
        {\bf Polarization of nuclear spin by repeated initialization.}
        The associated protocol is explained in Extended Data Fig.~\ref*{fig:extended_pulse_diagram}.
        The plot shows the degree of nuclear spin polarization $\exvalue{\Iz}$ versus the number of repetitions of the initialization protocol.
        We measured $\exvalue{\Iz}$ using spin tomography \cite{taminiau14}.
        Different colors represent different angles for the conditional X-rotations, varied from $\pi/2$ (green dots) to $0.125(\pi/2)$ (red dots).
        The plot demonstrates that even for incomplete X-rotations, polarization transfer from the NV center to the nuclear spin can still be achieved.
        This is relevant for a very weakly coupled nuclear spin, where the electron coherence time is too short to perform full $\pi/2$ X-rotations.}
    \label{fig:repetitive_initialization}
\end{figure*}

\begin{figure*}[h!]
    \centering
    \includegraphics[width=0.8\textwidth]{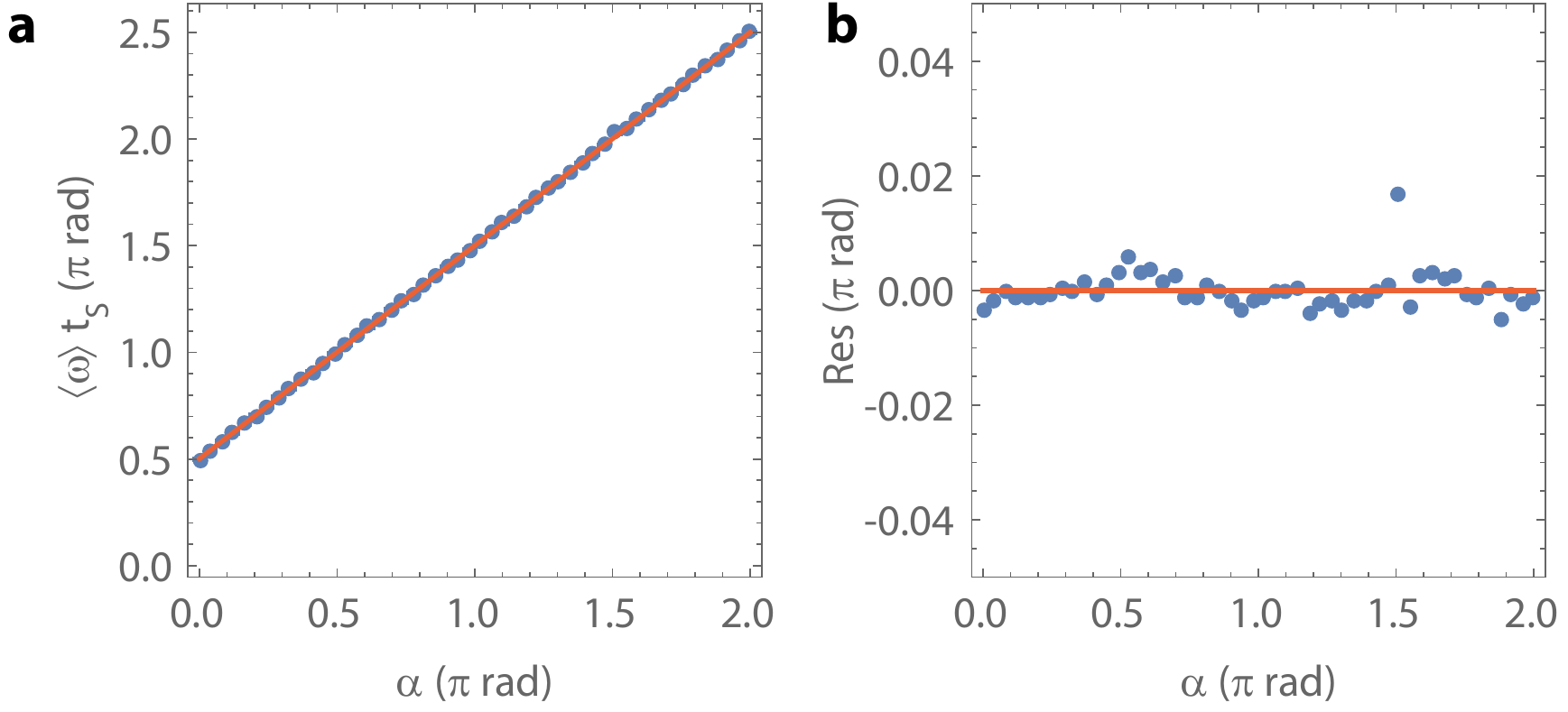}
    \caption{
		\linespread{\captionlinespread}\selectfont{}
        {\bf Statistical analysis of Fig. 3a, demonstrating that the precession frequency is not modified by sufficiently weak measurements.}
        {\bf a}, Signal frequency extracted from Lorentzian fits to each spectrum in Fig. 3a. We fit a linear function $ax+b$ and find $a=0.999305 \pm 0.000789$, $b=0.002171 \pm 0.000916$. A $\chi^{2}$ test yields $\chi^{2} = 0.000325$ and a corresponding $\text{p-val}=1.0$ according to the $\chi^2$ distribution for $k=50$ measurement points.
				{\bf b}, Residuals for the linear fit in (a). Frequency synchronization is absent in this plot because of the weak measurement strength ($\beta\sim 8^\circ$).
    }
    \label{fig:linearfreqfit_fig2res}
\end{figure*}

\begin{figure*}[h!]
    \centering
    \includegraphics[width=0.65\textwidth]{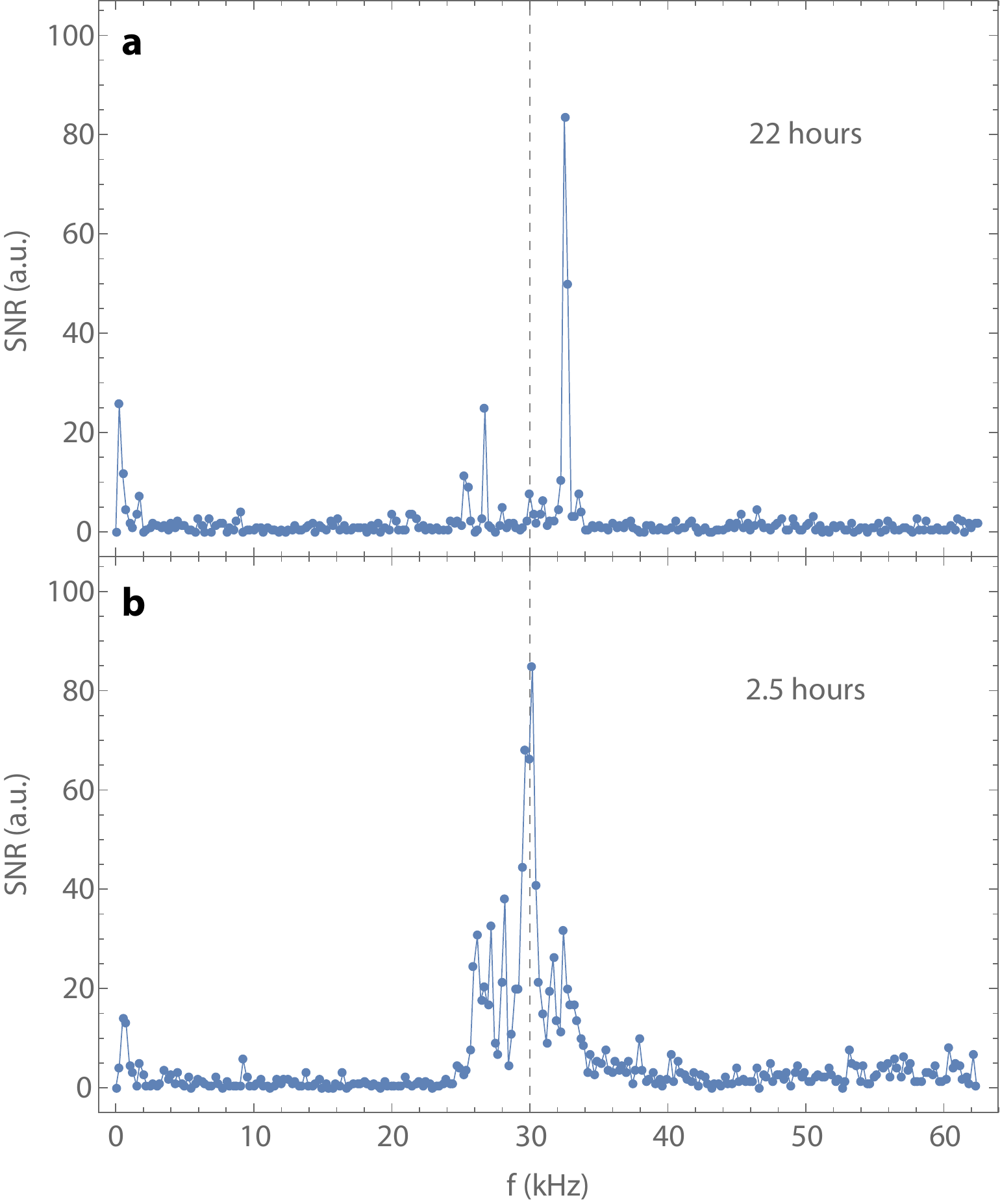}
    \caption{
		\linespread{\captionlinespread}\selectfont{}
        {\bf Weak measurement vs. nuclear Ramsey spectroscopy.}
        {\bf a}, Nuclear Ramsey spectroscopy of a \C spin bath. The plot shows the normalized power spectrum of a $4\unit{ms}$ long time trace after 22h of signal integration. As described in Extended Data Fig.\ref{fig:extended_pulse_diagram}b, we first polarize the \C spin bath, apply a $\pi/2$ rf-pulse to initiate precession and finally perform a strong measurement at the end of a variable free evolution time, which we increase in steps of $\ts=8\unit{\us}$. We included a $\pi$ pulse on the electron spin halfway the free evolution time. We also tuned the strong measurement to maximize the signal dynamic range. The peaks in the power spectrum are associated with individual \C spins that are relatively strongly coupled to the NV sensor such that a strong measurement is possible.
				{\bf b}, Weak measurement spectroscopy of the same \C spin bath (blue). The plot again shows the normalized power spectrum of a $4\unit{ms}$ long time trace, this time integrated during 2.5h. The presence of many more peaks around the bare Larmor frequency (gray dashed line) highlights a feature of continuous weak measurements: strongly coupled spins rapidly dephase, allowing for weaker signals (otherwise immersed in a strong background) to be detected. The couplings of these nuclei can be estimated from the spectral shift with respect to the bare Larmor frequency. For fast optical readout and under our measurement sequence, the observed shifts correspond to $\apar/(4\pi)$. The y-axis in both plots indicates the SNR calculated by normalizing the power spectrum amplitude to the standard deviation of the noise baseline (portion of the power spectrum where no signals are present). Both measurements were performed under identical initialization, sampling time and readout parameters (see Supplementary Data 1).
    }
    \label{fig:wm_dd_comparsion}
\end{figure*}
%

\begin{figure*}[h!]
    \centering
    \includegraphics[width=0.9\textwidth]{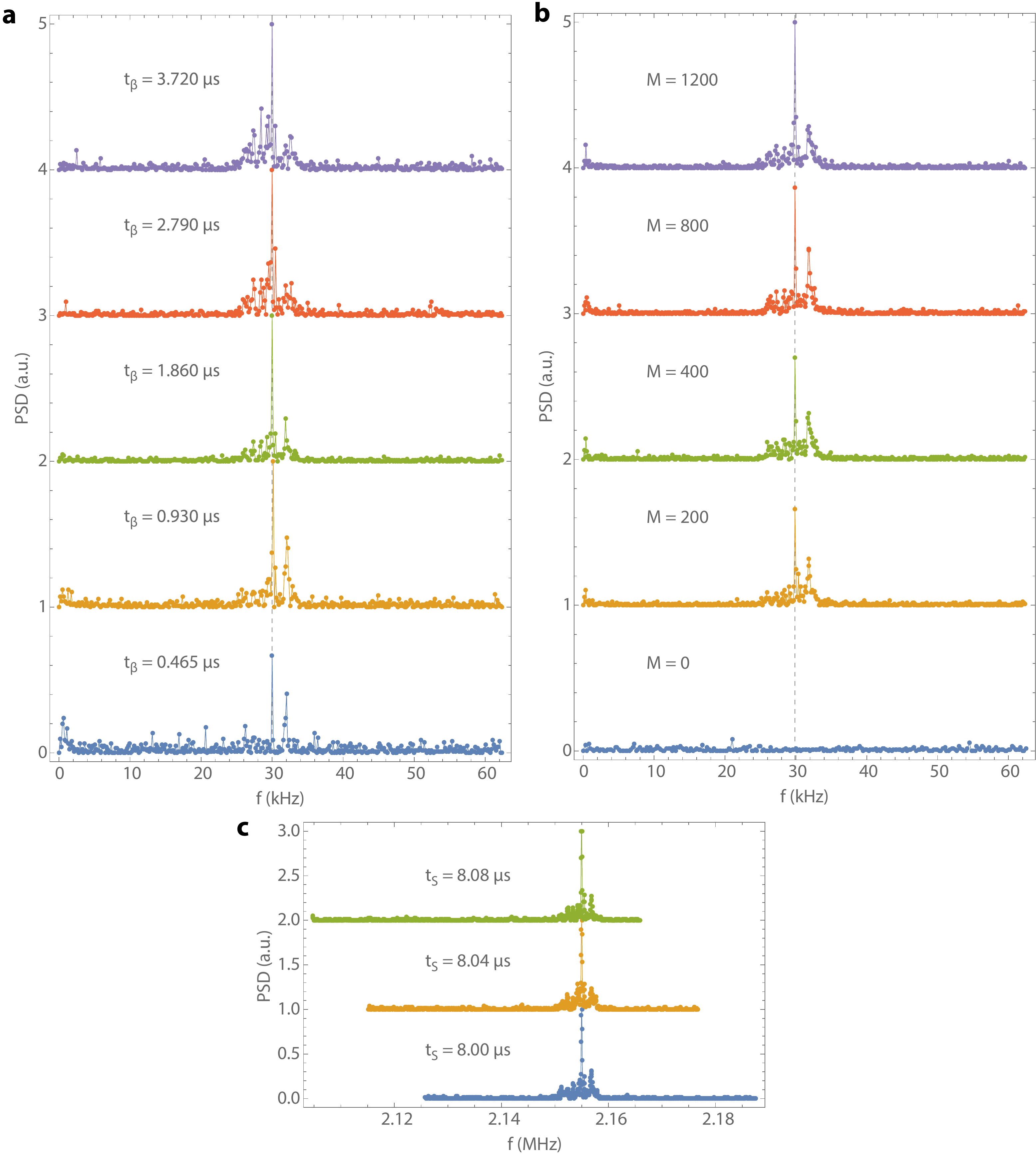}
    \caption{
		\linespread{\captionlinespread}\selectfont{}
		{\bf Dependence of weak measurement spectra on measurement strength, polarization time and sampling rate.}
        {\bf a}, Weak measurement power spectra for varying interaction time $\tbeta$, which defines the measurement strength $\beta = g\tbeta$.  Increasing $\tbeta$ (bottom to top) allows for more weakly coupled spins to be probed (peaks close to dashed line), while signals arising from nuclear spins with larger couplings $g$ become increasingly dephased (peaks far from dashed line).  Dashed line indicates the nuclear Zeeman frequency.
				{\bf b}, Weak measurement power spectra for different durations of nuclear polarization, increasing from bottom to top.  $M$ indicates the number of iterations of the NOVEL polarization transfer sequence (see Extended Data Fig. \ref{fig:extended_pulse_diagram}b).  The contact time for one iteration was $30\unit{\us}$. 
				{\bf c}, Weak measurement power spectra as a function of sampling time $\ts$.  Peak locations do not shift when varying the sampling time, indicating that no signals are folded due to aliasing.
				Spectra are vertically offset for clarity.
    }
    \label{fig:wm_check}
\end{figure*}

\begin{figure*}[h!]
    \centering
    \includegraphics[width=0.99\textwidth]{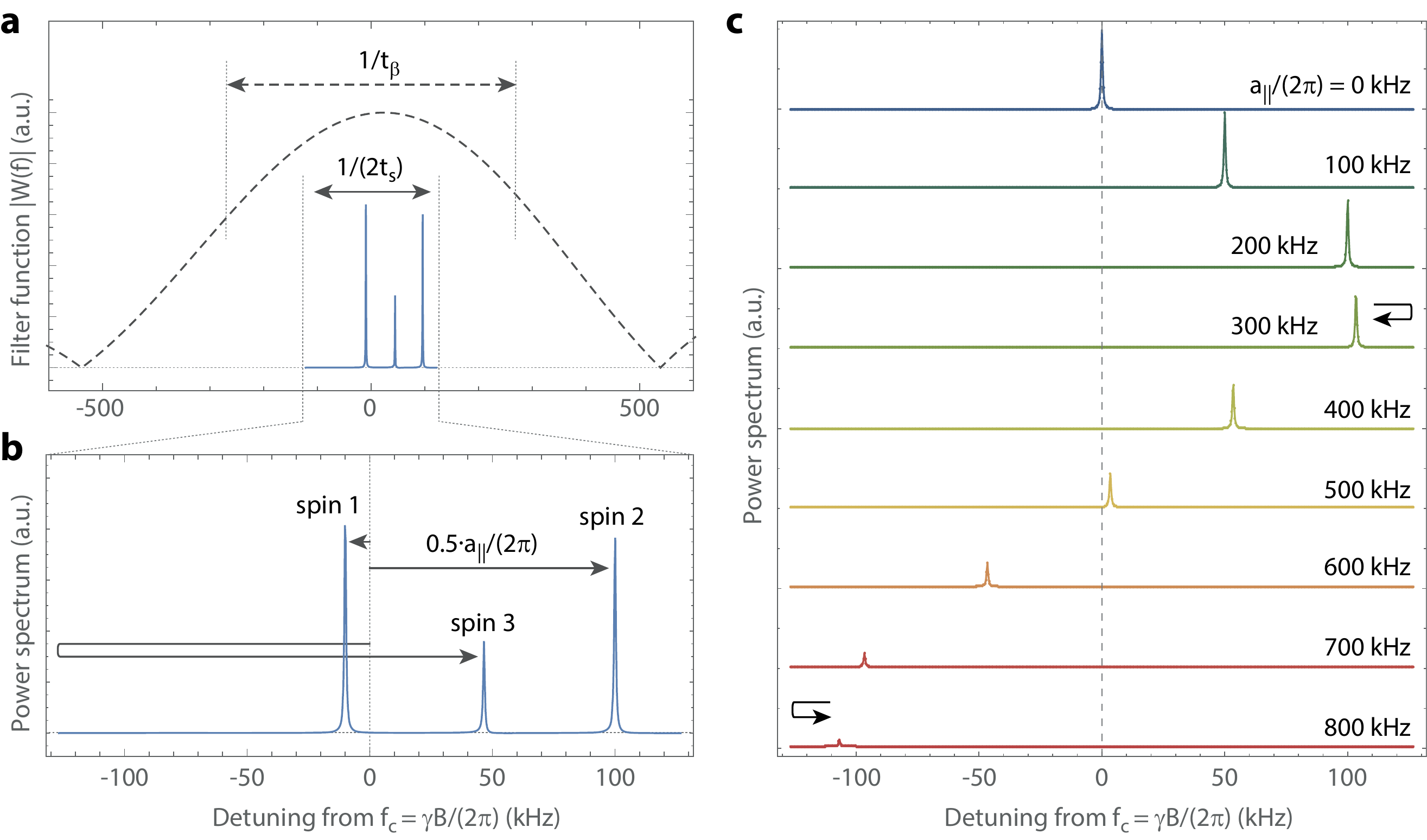}
    \caption{
		\linespread{\captionlinespread}\selectfont{}
		    {\bf Detection bandwidth of weak measurement spectroscopy.}
				{\bf a}, Calculated filter function \cite{biercuk10} of a DD sequence with $N=8$ pulses and $2\tau=232\unit{ns}$ interpulse delay [dashed line, Eq. (\ref{eq:ddfilter})].  The center frequency is $\fc=1/(4\tau)=2.154\unit{MHz}$, Ref. \onlinecite{degen17}, Eq. (68).  The nominal bandwidth of the DD sequence is $1/(\tbeta) = 1/(4N\tau) \approx 538\unit{kHz}$, Ref. \onlinecite{degen17}, Eqs. (68-69).  The Nyquist bandwidth is $1/(2\ts) \approx 254\unit{kHz}$, where we chose $\ts\approx\tbeta$ for the simulation. (In a real experiment, $\ts>\tbeta$ because of readout overhead).
				{\bf b}, Simulated weak measurement spectrum for three nuclear spins with parallel hyperfine parameters $\apar/(2\pi) = -20\unit{kHz}$ (spin 1), $\apar/(2\pi) = 200\unit{kHz}$ (spin 2) and $\apar/(2\pi) = -600\unit{kHz}$ (spin 3).  The transverse hyperfine parameter was $\aperp/(2\pi) = 5\unit{kHz}$ for all spins.  Note that the spectral shift for our scheme is $0.5\cdot \apar/(2\pi)$ (not $\apar/(2\pi)$, for further details see Extended Data Figure 1).  For spin 3, aliasing leads to folding of the signal peak back into the Nyquist bandwidth.  Simulations were implemented using density matrices.  The spin system included the central NV meter spin and three nuclear spins.  The NV center spin was implemented by a quasi spin-1/2 system consisting of the $m_S = {0, -1}$ sub-levels, and was simulated in the rotating frame of reference.
				(c) Simulated weak measurement spectra for a single spin whose parallel hyperfine parameter was increased from $\apar/(2\pi) = 0$ to $800\unit{kHz}$ in steps of $100\unit{kHz}$.  The transverse hyperfine parameter was $\aperp/(2\pi) = 5\unit{kHz}$.  The peak amplitudes clearly follow the profile of the filter function in (a), demonstrating that the detection bandwidth of weak measurement spectroscopy is determined by the wide DD filter function.  Arrows indicate back-folding of the peak due to aliasing. Note that for our specific experimental implementation, the detection of nuclear spins with strong $\apar$ couplings becomes difficult, due to inhomogeneous broadening caused by a residual hyperfine interaction during optical readout (see Methods).
    }
    \label{fig:wm_bw_nspins}
\end{figure*}

\begin{figure*}[h!]
    \centering
    \includegraphics[width=0.80\textwidth]{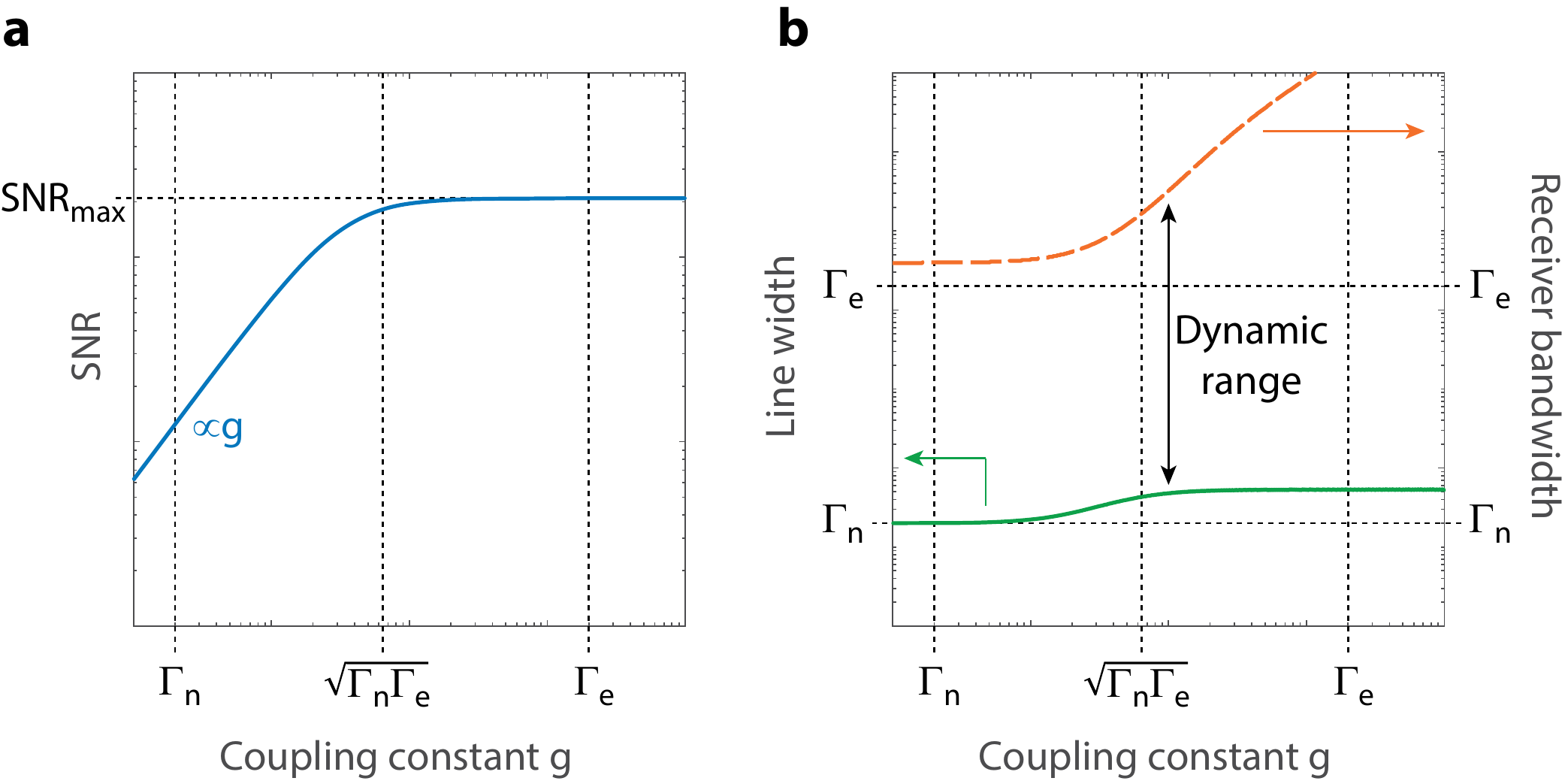}
    \caption{
		\linespread{\captionlinespread}\selectfont{}
		{\bf Qualitative scaling of signal-to-noise ratio, spectral resolution, and receiver bandwidth with coupling $g$}.
        {\bf a}, Log-log plot of the signal-to-noise ratio (SNR) per unit time as a function of the coupling parameter $g$.  $\Ge = (\Te)^{-1}$ is the decoherence rate of the electronic sensor spin and $\Gn = (\Tn)^{-1}$ is the dephasing rate of the nuclear spin, and we assume $\Ge\gg\Gn$.  Plotted curve is based on Supplementary Text 2.
        {\bf b}, Log-log plot of spectral line width (solid green curve) and receiver bandwidth (dashed red curve).  For weak measurement spectroscopy, the line width is approximately $\Gn$ (Supplementary Text 2).  The receiver bandwidth is $\tbeta^{-1}$ due to the DD filter function.  The frequency dynamic range of the measurement is given by the factor $\Gn\tbeta$ and can be very large, which is important for NMR spectroscopy applications.
    }
    \label{fig:sensitivity_comparison}
\end{figure*}

\end{document}